\documentclass[twocolumn,showpacs,preprintnumbers,amsmath,amssymb,prl]{revtex4-1}
\usepackage{graphicx}
\usepackage{dcolumn}
\usepackage{bm}
\usepackage{subfigure}
\usepackage{color}

\graphicspath{{./}{Figures//}}

\newcommand{\RAA}{\AA$^{-1}$}
\def\iat{Ir$_{1-x}$A$_x$Te$_2$}
\def\irt{Ir$_{1-x}$Rh$_x$Te$_2$}
\def\irte{Ir$_{0.8}$Rh$_{0.2}$Te$_2$}
\def\ipt{Ir$_{1-x}$Pt$_x$Te$_2$}
\def\ipte{Ir$_{0.95}$Pt$_{0.05}$Te$_2$}
\def\iprt{Ir$_{1-x}$(Pt,Rh)$_x$Te$_2$}
\def\ite2{IrTe$_2$}
\def\irtetwo{IrTe$_2$}
\def\cis{CuIr$_2$S$_4$}
\def\cics{Cu(Ir$_{1-x}$Cr$_x$)$_2$S$_4$}

\def\irtetwo{IrTe$_{2}$}
\def\t2g{$t2g$}

\def\p3barm1{P$\overline 1$}
\def\ponebar{P$\overline 1$}

\def\p-3m1{P$\overline 3$m1}
\def\i41amd{I4$_{1}$/amd}

\begin{document}
%
%
\begin{abstract}
The compound \irtetwo\ is known to exhibit a transition to a modulated state featuring Ir-Ir dimers, with large associated atomic
displacements. Partial substitution of Pt or Rh for Ir destabilizes the modulated structure and induces superconductivity. It has
been proposed that quantum critical dimer fluctuations might be associated with the superconductivity. Here we test for such local
dimer correlations and demonstrate their absence. X-ray pair distribution function approach reveals that the local structure of
\ipte\ and \irte\ dichalcogenide superconductors with compositions just past the dimer/superconductor boundary is explained well by
a dimer-free model down to 10~K, ruling out the possibility of there being nanoscale dimer fluctuations in this regime. This is inconsistent with
the proposed quantum-critical-point-like interplay of the dimer state and superconductivity, and precludes scenarios for dimer fluctuations
mediated superconducting pairing.
\end{abstract}

\title{Absence of Local Fluctuating Dimers in Superconducting \iprt}

    \author{Runze~Yu,$^{1,\dag}$ S. Banerjee,$^{2}$ {H.~C.} ~Lei,$^{1,\dag\dag}$ Ryan Sinclair,$^{3}$  M. Abeykoon,$^{4}$ {H.~D.} Zhou,$^{3}$  C. Petrovic,$^{1}$ Z. Guguchia,$^{1,5}$ and E. S. Bozin$^{1,*}$}
    \affiliation{$^{1}$Condensed Matter Physics and Materials Science Department, Brookhaven National Laboratory, Upton, NY~11973, USA}
    \altaffiliation{bozin@bnl.gov\\
                    $^\dag$Present address: Institute of Physics, Chinese Academy of Science, Beijing, 100190, Peoples Republic of China\\
                    $^{\dag\dag}$ Present address: Department of Physics, Renmin University, Beijing 100872, Peoples Republic of China}
    \affiliation{$^{2}$Department of Applied Physics and Applied Mathematics, Columbia University, New York, NY~10027, USA}
    \affiliation{$^{3}$Department of Physics and Astronomy, University of Tennessee, Knoxville, Tennessee 37996, USA}
    \affiliation{$^{4}$Photon Sciences Division, Brookhaven National Laboratory, Upton, NY 11973, USA}
    \affiliation{$^{5}$Department of Physics, Columbia University, New York, NY~10027, USA}

\date{\today}
\maketitle

%
%
%
Unconventional superconductivity (SC) often emerges in the proximity of symmetry breaking
electronic and magnetic orders upon their destabilization by chemical modifications,
external pressure and fields, as seen in a diverse variety of quantum systems~\cite{basov;np11,saito;chrec11,jiao;pnas15}.
The pairing mechanism remains elusive~\cite{norma;s11}, in part because
the role of fluctuations of adjacent ordered states and their ubiquity are not fully established
and understood~\cite{lee;rmp06,chang;np12,comin;s14}.
Studying such fluctuations is quite challenging~\cite{kivel;rmp03}, one of the reasons being
the lack of the long range coherence~\cite{billi;s07}. When broken symmetry states, for example
electronic states involving 5$d$ manifolds in \cis~\cite{radae;n02,khoms;prl05} and \irtetwo~\cite{pascu;prl14,toriy;jpsj14}
where two Ir$^{4+}$ S=1/2 bind into spinless spatially ordered dimers, are coupled to the lattice, footprints of their fluctuations
become evident in the local atomic structure and can be studied indirectly using a local structural probe~\cite{lee;n06,bozin;prl07}
such as the atomic pair distribution function (PDF) analysis of powder diffraction data~\cite{billi;jssc08,egami;b;utbp12}.
Here we use x-ray PDF to probe the existence or absence of Ir$^{4+}$-Ir$^{4+}$ dimer fluctuations in
doped \irtetwo\ superconductor, which yields information essential for bona fide considerations
of dimer/SC entanglement in this system.

Trigonal metallic iridium ditelluride, \irtetwo, has garnered significant attention over the past several
years following the discovery of bulk superconductivity ($T_{c}\sim$3~K) in its intercalated and
substituted variants IrTe$_{2}$:Pd~\cite{yang;prl12}, Ir$_{1-x}$Pt$_{x}$Te$_{2}$~\cite{pyon;jpsj12},
Cu$_{x}$IrTe$_{2}$~\cite{kamit;prb13}, and  Ir$_{1-x}$Rh$_{x}$Te$_{2}$~\cite{kudo;jpsj13}.
Interestingly, the appearance of SC also follows the suppression of a long range ordered electronic
state, in this case associated with charge disproportionation enabled Ir$^{4+}$-Ir$^{4+}$ dimerization~\cite{pascu;prl14,toriy;jpsj14}
established in \irtetwo\ at its symmetry lowering structural transition ($T_{s}\sim$250~K)~\cite{matsu;jltp99}.
This results in familiar domelike phase diagrams, akin to those of high temperature SCs and recently discovered
Cu$_{x}$TiSe$_{2}$~\cite{moros;np06}, 1T-TaS$_{2}$~\cite{sipos;nm08}, 1T-TiSe$_{2}$~\cite{kusma;prl09},
T$_{d}$-MoTe$_{2}$~\cite{guguc;nc17}, ZrTe$_{3-x}$Se$_{x}$~\cite{zhu;sr16}, and 2H-TaSe$_{2-x}$S$_{x}$~\cite{li;npjqm17}
transition metal dichalcogenide superconductors, where destabilization of the charge density wave (CDW) order
leads to SC. Importantly, in Cu$_{x}$TiSe$_{2}$ quantum criticality associated with fluctuations of CDW order
has been considered in relation to SC pairing~\cite{barat;prl08,castr;prl01}. A perceived analogy with these
systems prompted a hypothesis of quantum critical point (QCP) like interplay of SC and dimerization
in \irtetwo\ derivates~\cite{yang;prl12}, and speculations about dimer fluctuation mediated superconductivity~\cite{pyon;jpsj12,dai;prb14}.

\begin{figure}[tb]
\includegraphics[width=0.85\columnwidth]{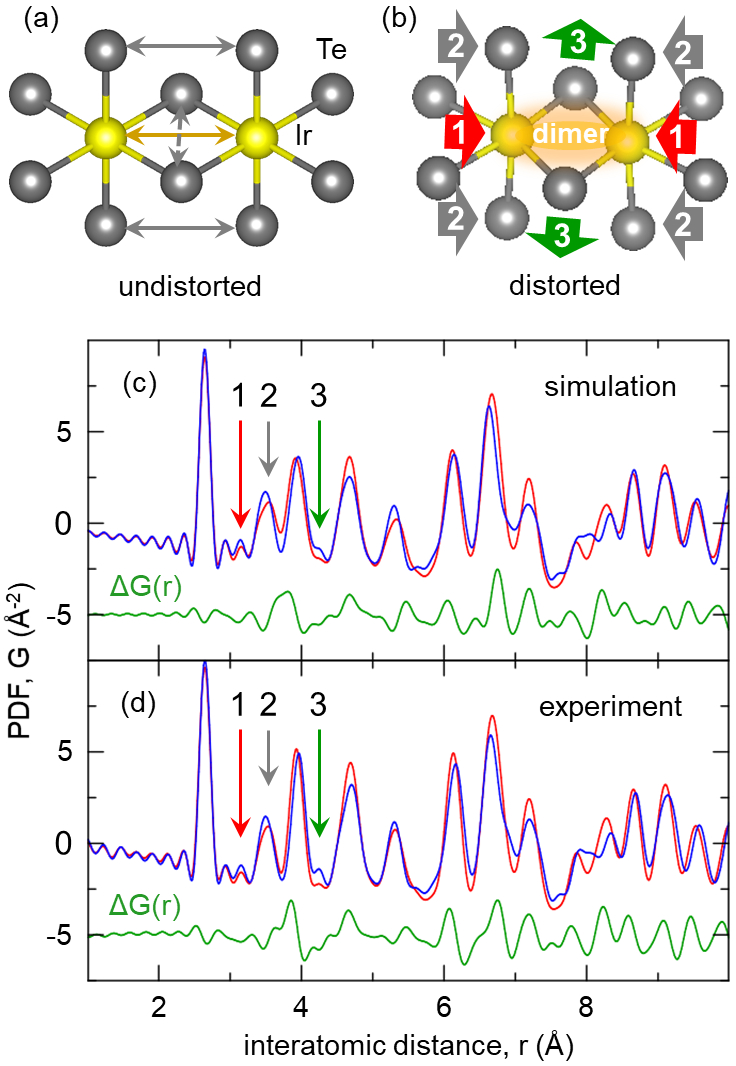}
\caption{\label{fig:structure} (Color online) Sketch of the local atomic environments in \irtetwo\ for the (a) undistorted
high temperature structure (trigonal \p-3m1), and
(b) distorted low temperature structure (triclinic \ponebar) featuring Ir-Ir and Te-Te dimers.
Dimerization results in dramatic distortions of associated interatomic
distances relative to the high temperature structure, as indicated by block arrows and described in the text.
Comparison of PDFs (c) calculated from trigonal (red line) and
triclinic (blue line) models, and (d) measured at 275~K (red line) and at 220~K (blue line). Enumerated vertical arrows in (c) and (d)
mark features associated with these distortions. $\Delta G(r)$ is the difference, offset for clarity.
}
\end{figure}

The importance of the \irtetwo\ lattice in facilitating the long range dimer order is
well documented~\cite{oh;prl13,kim;prl15,ivash;sr17}, with signatures of the dimer
state found in a remarkable reduction of intradimer Ir-Ir (0.8~\AA)
and associated Te-Te (0.5~\AA) distances~\cite{pascu;prl14}, as illustrated in Fig.~\ref{fig:structure}(a), (b).
Despite this, and the importance of verifying the dimer fluctuations hypothesis, the utilization of
experimental probes sensitive to presence/absence of local distortions has been surprisingly scarce.
Existing reports based on extended x-ray absorption fine structure (EXAFS) spectroscopy
focus on parent \irtetwo\ under ambient~\cite{josep;prb13} and high pressure~\cite{paris;prb16}
conditions. The ambient study argues for persistence of \emph{local} Ir dimers
in the high temperature regime where the structure is undistorted trigonal on average~\cite{josep;prb13}.
Whilst this, if true, could hint at the presence of fluctuating dimers also in the superconducting compositions,
experimental validation is still lacking.
%
%

Here we employ the PDF approach on superconducting compositions of two different \iat\ families
just across the dimer/SC boundary to explore for the first time the existence of local dimer fluctuations.
The PDF sensitivity to the presence of Ir-Ir dimers irrespective of the character of
their ordering has been demonstrated in \cis~\cite{radae;n02,bozin;prl11} and \cics~\cite{bozin;sr14}
spinels, where similar dimerization takes place on the Ir pyrochlore sublattice. When present, local dimers are clearly
evident in the PDF of \irtetwo\ due to the large change in the Ir-Ir and Te-Te interatomic distances associated
with them. Here we provide conclusive evidence that the dimers are absent in \ipte\ and \irte\
down to 10~K. This unambiguously rules out the popular hypothesis of quantum dimer fluctuations in this regime and that such
fluctuations play a role in SC pairing. Moreover, PDF finds no evidence for dimer fluctuations in
\irtetwo\ at $T>T_{s}$, in stark contrast to previous EXAFS report~\cite{josep;prb13}.

Polycrystalline  samples of \irtetwo, \ipte, and \irte\ were synthesized using standard
solid-state protocols, and were found to be single phase based on x-ray powder diffraction~\cite{cao;prb13,lazar;prb14}.
Total scattering PDF experiments were performed at the 28-ID-2 beam line at the National Synchrotron Light Source II
at Brookhaven National Laboratory, with 67.7~keV x-rays using the rapid acquisition mode with 60~s exposure/dataset~\cite{chupa;jac03}. The setup
utilized {\it Perkin-Elmer} area detector and {\it Cryoindustries of America} cryostat for data collection between 10~K and
300~K on warming.
The raw 2D diffraction data were integrated and converted to intensity versus $Q$ using the software
{\sc fit2d}~\cite{hamme;hpr96}, where $Q$ is the magnitude of the scattering vector. Data reduction to measured
total scattering structure functions, $F(Q)$, and their successive Sine Fourier transform up to a momentum transfer
of $Q_{max}= 25$~\RAA\ to obtain experimental PDFs, $G(r)$, were carried out using the {\sc pdfgetx3}~\cite{juhas;jac13}
program.  Models with \p-3m1\ and \ponebar\ symmetry were used to describe nondimerized (Fig.~\ref{fig:structure}(a))
and dimerized (Fig.~\ref{fig:structure}(b)) structures, respectively, using the {\sc pdfgui} suite~\cite{farro;jpcm07}.

We begin by establishing qualitatively the sensitivity of our PDF data to the presence of dimers and concomitant
structural distortions in \irtetwo. In the high temperature phase above $T_{s}$ all Ir atoms are in
identical Te$_{6}$ octahedral environments displaying an edge-shared topology, Fig.~\ref{fig:structure}(a),
constituting trigonal symmetry average structure~\cite{hocki;jpc60}.
In the low temperature phase just below $T_{s}$, where the dimer patterns with a stripe morphology corresponding to {\bf q$_{0}$} = 1/5(1, 0, 1) ordering
are established~\cite{yang;prl12,oh;prl13}, Ir atoms subject to dimerization sit in distorted Te$_{6}$ octahedral environments, Fig.~\ref{fig:structure}(b),
and the average symmetry lowers to triclinic~\cite{pascu;prl14}. Pairs of dimerization-affected IrTe$_{6}$
octahedra exhibit dramatic structural rearrangements: Ir-Ir and Te-Te dimer distances reduce by
0.8~\AA\ and 0.5~\AA\ respectively, while the lateral Te-Te distance (common edge) elongates by 0.3~\AA~\cite{pascu;prl14}.
The distortions are depicted by enumerated block arrows in Fig.~\ref{fig:structure}(b). Importantly, only $\sim$ 6~\% of
all nearest neighbor Ir-Ir distances on triangular Ir planes of \irtetwo\ dimerize, in contrast to \cis\ where
the fraction of dimerized Ir contacts is about 5 times larger~\cite{radae;n02}.

%
\begin{figure}[tb]
\includegraphics[width=0.85\columnwidth]{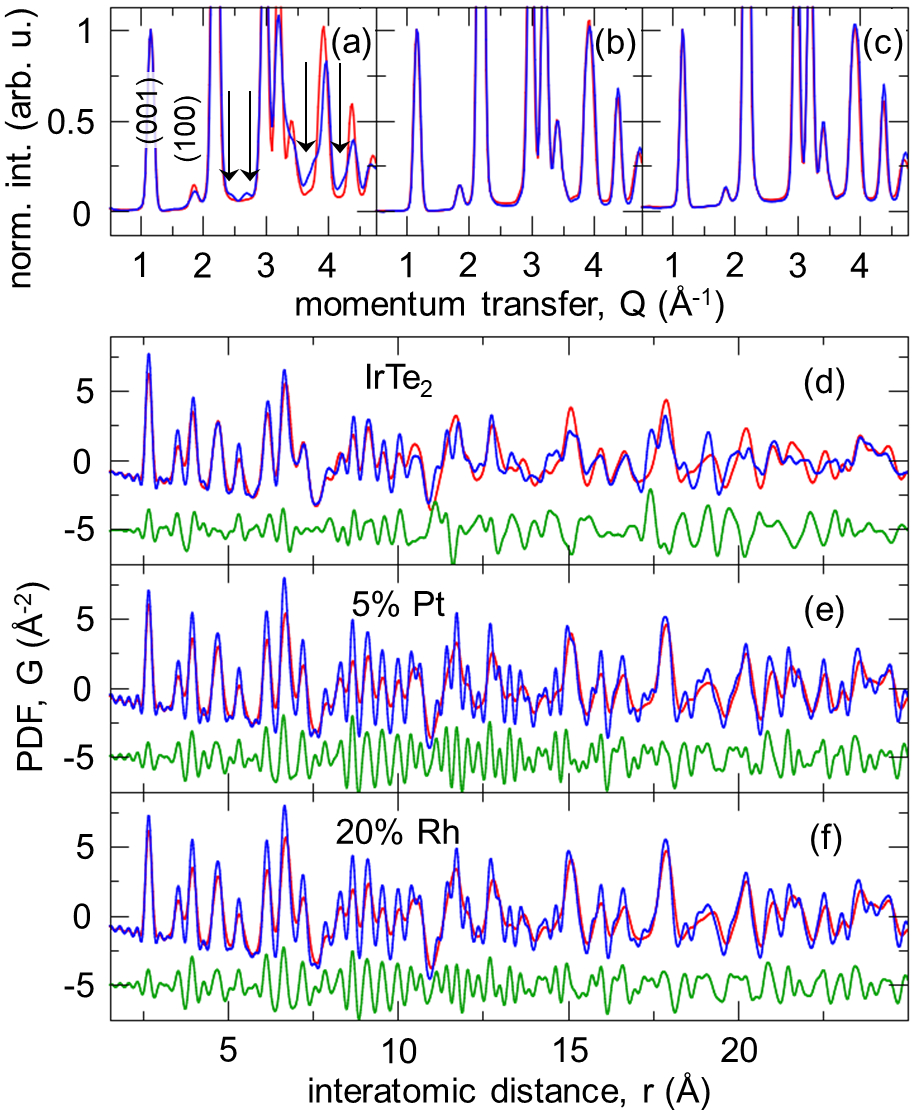}
\caption{\label{fig:data} (Color online) Azimuthally integrated 2D diffraction patterns of \iprt\ for 300~K
(red line) and 10~K (blue line) over a narrow range of momentum transfer, $Q$, for (a) x=0, (b) x=0.05 Pt, and
(c) x=0.2 Rh. All patterns are normalized by the intensity of (001) reflection (\p-3m1\ indexing).
Vertical arrows in (a) indicate superlattice reflections observed in 10~K data. Corresponding PDFs are
compared in (d), (e), and (f), respectively, with differences shown underneath and offset for clarity.}
\end{figure}
%
We simulated PDF patterns for the average crystal structures for $T>T_{s}$ (trigonal) and  $T<T_{s}$ (triclinic)
using parameters from single crystal x-ray diffraction~\cite{pascu;prl14}.
These are shown in Fig.~\ref{fig:structure}(c) as red and blue profiles, respectively, with their difference
plotted underneath. Changes in the interatomic distance distribution arising from dimerization as
seen by PDF are marked by enumerated vertical arrows. Examination of the high and low temperature profiles
reveals a redistribution of intensity in PDF peaks centered around 3.5~\AA\ (Te-Te) and 3.9~\AA\ (lattice repeat
distance), whereas new peaks appear at around 3.1~\AA\ (Ir-Ir dimer), 3.4~\AA\ (Te-Te dimer), and 4.2~\AA\ (common Te-Te edge).
It is evident that the Ir-Ir dimer signal at 3.1~\AA\ is rather weak, as compared to that
observed in \cis\ spinel~\cite{bozin;prl11}, and is barely visible above the parapet of termination ripples caused
by the finite range of the Fourier transform. This comes about due to different dimer densities in the two materials.
However, this analysis shows that, despite this relatively weaker signal, the PDF is still sensitive to
the presence or absence of local dimers.

Experimental PDFs of \irtetwo\ for temperatures straddling $T_{s}$ are compared in Fig.~\ref{fig:structure}(d),
where the 275~K (red profile, $T>T_{s}$) and the 220~K (blue profile, $T<T_{s}$) data and their difference are displayed.
A qualitative assessment readily demonstrates that all dimerization features described above and highlighted in
the calculated PDFs, which contain the impact of the dimers, are well reproduced in the experimental PDF data. This clearly
establishes the PDF sensitivity to dimer structural signatures and their detectability in our data.
Comparisons in Figs.~\ref{fig:structure} (c) and (d) also indicate that the dimers \emph{disappear} in the local structure
above $T_{s}$, as is further confirmed by explicit
modeling that we discuss later. Notably, proposed order-disorder scenario for the dimerization transition~\cite{josep;prb13} is at odds
with this observation. The first-order nature of the transition~\cite{matsu;jltp99,oh;prl13} also argues against the persistence of local fluctuating dimers
above $T_{s}$.

%
\begin{figure}[tb]
\includegraphics[width=0.9\columnwidth]{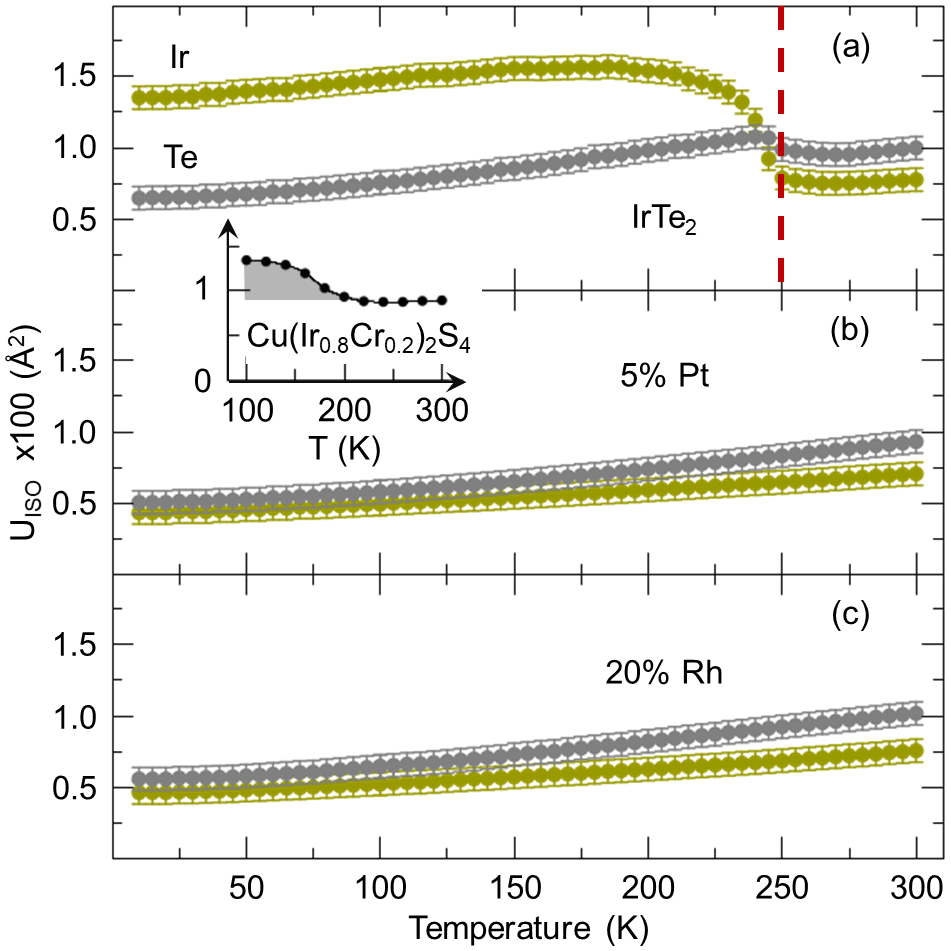}
\caption{\label{fig:adps} (Color online) Temperature dependence of the isotropic ADPs of Ir (olive symbols) and Te (gray symbols)
in \iprt\ obtained from \p-3m1\ model fits to (a) x=0, (b) x=0.05 Pt, and (c) x=0.2 Rh sample data over the 10~K--300~K temperature range.
Inset: detection of onset of dimer fluctuations from temperature dependent Ir-ADP in 20\% Cr-doped \cis, where Ir-Ir dimerization sets in
below 200~K on a nanometer length-scale \emph{only}, while the long range dimer order is absent at all temperatures~\cite{bozin;sr14}.
Such behavior is \emph{not} observed in (b) and (c) for superconducting \iprt.
}
\end{figure}
%
Samples with SC compositions display qualitatively different behavior on all lengthscales accessible by
our measurements. When the average \irtetwo\ symmetry is lowered
and the long range dimer order is established, superlattice reflections appear in the
integrated diffraction patterns, as seen in Fig.~\ref{fig:data}(a) where 10~K and
300~K are compared. In contrast, no such features are observed
in \ipte\ (Fig.~\ref{fig:data}(b)) and \irte\ (Fig.~\ref{fig:data}(c)) data at any temperature,
consistent with the average symmetry remaining trigonal down to 10~K and no long
range dimer order, as expected from the monotonic temperature variation of susceptibility
and electrical resistivity~\cite{pyon;jpsj12,kudo;jpsj13}.
Importantly, the dimers are also not observed at low temperature on intermediate and short lengthscales
probed by the PDF. When symmetry lowering occurs, this causes redistribution of PDF intensities and overall broadening of
the PDF patterns due to the appearance of new interatomic distances. Conversely, temperature lowering sharpens the PDF features
as a consequence of decreasing the amplitudes of thermal vibrations~\cite{egami;b;utbp12}. Both effects
are present in \irtetwo\ PDFs, Fig.~\ref{fig:data}(d), where the 300~K profile is observably sharper
than that of 10~K at intermediate $r$, and their difference reveals a change corresponding
to a superposition of these two opposite effects. Figs.~\ref{fig:data} (e) and (f) show
300~K and 10~K data for superconducting samples. Whilst there are also dramatic changes evidenced in
the respective difference curves, this is qualitatively different from what is seen in \irtetwo.

It can be shown through a semiempirical scaling procedure that high and low temperature PDFs of doped
samples can be successfully morphed into each other, whereas such a procedure fails in the case of \irtetwo.
This arguably indicates that the changes in the SC samples are likely caused solely by
thermal effects, without symmetry lowering, whereas the actual symmetry breaking is needed to
explain the changes in the parent system (see Supplemental Material for details). To further corroborate this,
an undistorted trigonal model was refined in the $r$-space against the PDF data in 10~K--300~K range for all samples, and
the obtained atomic displacement parameters (ADPs) monitored, Fig.~\ref{fig:adps}. While in \irtetwo\,
both ADPs of Ir and Te initially drop linearly with temperature, Fig.~\ref{fig:adps}(a),
they exhibit an abrupt jump at the onset of the dimerization transition, denoted by vertical dashed red
line in the figure. This nominally implies ''disorder'', but actually reflects the inadequacy of the
trigonal model to explain the symmetry breaking and underlying dimerization encoded in the data.
In contrast, no such jumps are observed for \ipte\ (Fig.~\ref{fig:adps}(b)) and \irte\ (Fig.~\ref{fig:adps}(c))
in the entire temperature range studied. It is important to realize that jumps in ADPs are not only observed
across long range symmetry breaking transitions, but also in cases when there is only a local structure
change in the absence of any macroscopic transitions. The inset in Fig.~\ref{fig:adps} exemplifies such a
situation seen in 20\% Cr-doped \cis, where \emph{local} Ir-Ir dimerization sets in just below 200~K,
in the absence of long range dimer order at any temperature in that system~\cite{bozin;sr14}.
This demonstrates that there is neither average nor local symmetry lowering in \ipte\ and \irte\
down to 10~K.

%
\begin{figure}[tb]
\includegraphics[width=0.85\columnwidth]{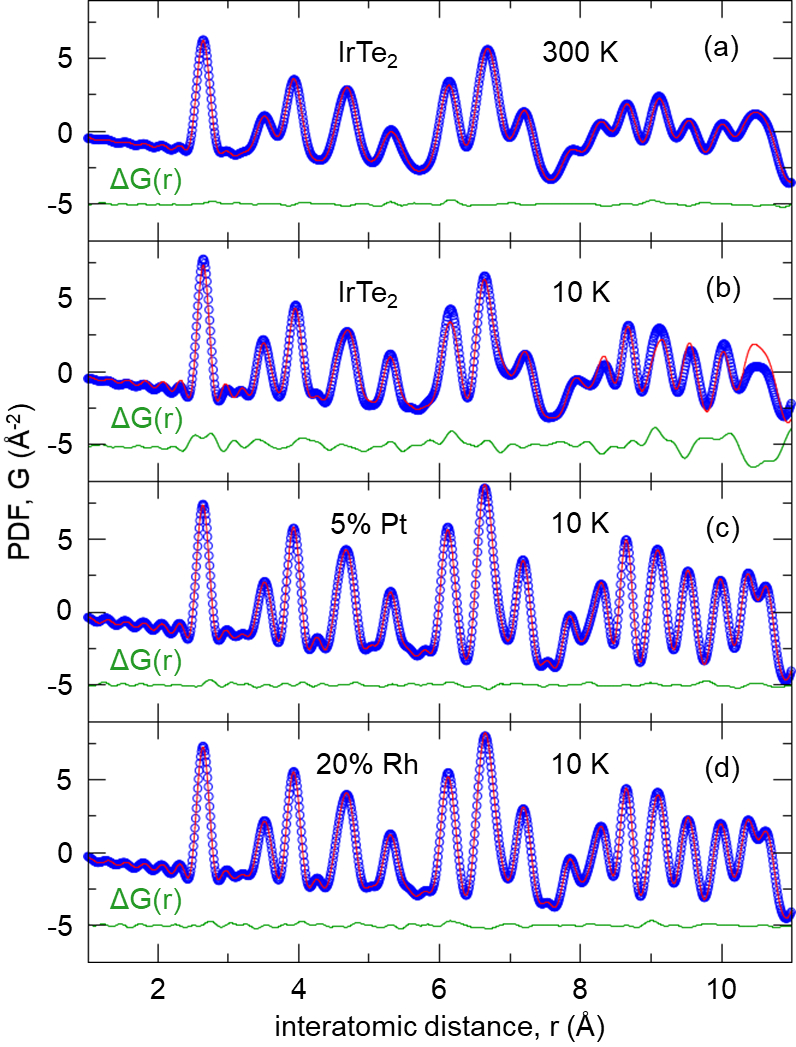}
\caption{\label{fig:comparison} (Color online) Fits of trigonal structure model (solid red lines) to experimental PDF data (open blue symbols) of
\irtetwo\ sample at (a) 300~K and (b) 10~K, 5~\% Pt substituted sample at 10~K (c), and 20~\% Rh substituted sample at 10~K (d). The difference between
the data and the model, $\Delta G(r)$, plotted below, is offset for clarity. See text for details.
}
\end{figure}
%
Finally, we consider the actual fits of the trigonal structure model to various PDF data over the 10 \AA\ range,
shown in Fig.~\ref{fig:comparison} as solid red lines and open blue symbols, respectively. This
undistorted dimer-free model explains the \irtetwo\ 300~K data exceptionally well, Fig.~\ref{fig:comparison}(a),
as evident from the flat difference curve and a low fit residual value of $r_{w}\sim$4\%.
The same model fails to explain the \irtetwo\ data at 10~K ($r_{w}\sim$15\%), in Fig.~\ref{fig:comparison}(b), as expected, given
that at that temperature long range dimer order is well established and that the attempted dimer-free model
is strictly inadequate. Importantly, this failed fit charts substantial misfits in the difference
curve that would reveal the presence of dimers in the data when they are confronted with a dimer-free model.
Figs.~\ref{fig:comparison}(c) and (d) show the results of such a fitting attempt carried out on 10~K \ipte\ and
\irte\ data, respectively. Not only do the corresponding difference curves not display the features observed in
Fig.~\ref{fig:comparison}(b), but the fits of the trigonal model in fact agree rather well with the data ($r_{w}\sim$5\%).
The local structure at 10~K for these compositions can be well explained by a dimer-free model.
This quantitative analysis validates the aforementioned qualitative conclusions about the absence of dimer
fluctuations in the high temperature phase of the parent system as well as at 10~K in superconducting compositions
just past the dimer/SC boundary. The transition in \irtetwo\ is argued to originate from a uniform lattice deformation
combined with charge ordering and subsequent Ir dimerization~\cite{kim;prl15}. According to a recent high pressure study
of \ipt, the structural transition triggers charge ordering and dimerization~\cite{ivash;sr17}. This implies that fluctuations
associated with the putative QCP are expected to appear not only in the electronic, but also in the structural channel. However,
the PDF results presented here do not show any evidence that would support this picture.

In conclusion, by using state of the art x-ray total scattering based PDF approach we establish the first direct evidence
for the absence of local dimer fluctuations in the phase diagrams of \ipt\ and \irt\ beyond the dimer/superconductor
phase boundary. The dimer fluctuations are also absent in the parent \irtetwo\ in the temperature regime above the structural
phase transition. These results imply that dimer fluctuations are not a relevant part of the phase diagram of \irtetwo\
based systems and thus their role in the superconducting pairing is implausible. The results provide important new
constraints for theoretical considerations of the complex interplay
between superconductivity and other electronic orders in this class of materials.

\begin{acknowledgments}
Work at Brookhaven National Laboratory was supported by US DOE, Office of Science, Office of Basic Energy
Sciences (DOE-BES) under contract DE-SC00112704. R. Sinclair and H. D. Zhou acknowledge support from
NSF-DMR-1350002. We are grateful to John Tranquada, Simon Billinge, Ian Robinson, and Alexei Tsvelik
for fruitful discussions and critical comments.
\end{acknowledgments}



\begin{thebibliography}{45}
\expandafter\ifx\csname natexlab\endcsname\relax\def\natexlab#1{#1}\fi
\expandafter\ifx\csname bibnamefont\endcsname\relax
  \def\bibnamefont#1{#1}\fi
\expandafter\ifx\csname bibfnamefont\endcsname\relax
  \def\bibfnamefont#1{#1}\fi
\expandafter\ifx\csname citenamefont\endcsname\relax
  \def\citenamefont#1{#1}\fi
\expandafter\ifx\csname url\endcsname\relax
  \def\url#1{\texttt{#1}}\fi
\expandafter\ifx\csname urlprefix\endcsname\relax\def\urlprefix{URL }\fi
\providecommand{\bibinfo}[2]{#2}
\providecommand{\eprint}[2][]{\url{#2}}

\bibitem[{\citenamefont{Basov and Chubukov}(2011)}]{basov;np11}
\bibinfo{author}{\bibfnamefont{D.~N.} \bibnamefont{Basov}} \bibnamefont{and}
  \bibinfo{author}{\bibfnamefont{A.~V.} \bibnamefont{Chubukov}},
  \bibinfo{journal}{Nature Phys.} \textbf{\bibinfo{volume}{7}},
  \bibinfo{pages}{272} (\bibinfo{year}{2011}).

\bibitem[{\citenamefont{Saito and Yoshida}(2011)}]{saito;chrec11}
\bibinfo{author}{\bibfnamefont{G.}~\bibnamefont{Saito}} \bibnamefont{and}
  \bibinfo{author}{\bibfnamefont{Y.}~\bibnamefont{Yoshida}},
  \bibinfo{journal}{Chem. Rec.} \textbf{\bibinfo{volume}{11}},
  \bibinfo{pages}{124} (\bibinfo{year}{2011}).

\bibitem[{\citenamefont{Jiao et~al.}(2015)\citenamefont{Jiao, Chen, Kohama,
  Graf, Bauer, Singleton, Zhu, Weng, Pang, Shang et~al.}}]{jiao;pnas15}
\bibinfo{author}{\bibfnamefont{L.}~\bibnamefont{Jiao}},
  \bibinfo{author}{\bibfnamefont{Y.}~\bibnamefont{Chen}},
  \bibinfo{author}{\bibfnamefont{Y.}~\bibnamefont{Kohama}},
  \bibinfo{author}{\bibfnamefont{D.}~\bibnamefont{Graf}},
  \bibinfo{author}{\bibfnamefont{E.}~\bibnamefont{Bauer}},
  \bibinfo{author}{\bibfnamefont{J.}~\bibnamefont{Singleton}},
  \bibinfo{author}{\bibfnamefont{J.}~\bibnamefont{Zhu}},
  \bibinfo{author}{\bibfnamefont{Z.}~\bibnamefont{Weng}},
  \bibinfo{author}{\bibfnamefont{G.}~\bibnamefont{Pang}},
  \bibinfo{author}{\bibfnamefont{T.}~\bibnamefont{Shang}},
  \bibnamefont{et~al.}, \bibinfo{journal}{Proc. Natl. Acad. Sci. USA}
  \textbf{\bibinfo{volume}{112}}, \bibinfo{pages}{673} (\bibinfo{year}{2015}).

\bibitem[{\citenamefont{Norman}(2011)}]{norma;s11}
\bibinfo{author}{\bibfnamefont{M.~R.} \bibnamefont{Norman}},
  \bibinfo{journal}{Science} \textbf{\bibinfo{volume}{332}},
  \bibinfo{pages}{196} (\bibinfo{year}{2011}).

\bibitem[{\citenamefont{Lee et~al.}(2006{\natexlab{a}})\citenamefont{Lee,
  Nagaosa, and Wen}}]{lee;rmp06}
\bibinfo{author}{\bibfnamefont{P.~A.} \bibnamefont{Lee}},
  \bibinfo{author}{\bibfnamefont{N.}~\bibnamefont{Nagaosa}}, \bibnamefont{and}
  \bibinfo{author}{\bibfnamefont{X.}~\bibnamefont{Wen}}, \bibinfo{journal}{Rev.
  Mod. Phys.} \textbf{\bibinfo{volume}{78}}, \bibinfo{pages}{17}
  (\bibinfo{year}{2006}{\natexlab{a}}).

\bibitem[{\citenamefont{Chang et~al.}(2010)\citenamefont{Chang, Blackburn,
  Holmes, Christensen, Larsen, Mesot, Liang, Bonn, Hardy, Watenphul
  et~al.}}]{chang;np12}
\bibinfo{author}{\bibfnamefont{J.}~\bibnamefont{Chang}},
  \bibinfo{author}{\bibfnamefont{E.}~\bibnamefont{Blackburn}},
  \bibinfo{author}{\bibfnamefont{A.~T.} \bibnamefont{Holmes}},
  \bibinfo{author}{\bibfnamefont{N.~B.} \bibnamefont{Christensen}},
  \bibinfo{author}{\bibfnamefont{J.}~\bibnamefont{Larsen}},
  \bibinfo{author}{\bibfnamefont{J.}~\bibnamefont{Mesot}},
  \bibinfo{author}{\bibfnamefont{R.}~\bibnamefont{Liang}},
  \bibinfo{author}{\bibfnamefont{D.~A.} \bibnamefont{Bonn}},
  \bibinfo{author}{\bibfnamefont{W.~N.} \bibnamefont{Hardy}},
  \bibinfo{author}{\bibfnamefont{A.}~\bibnamefont{Watenphul}},
  \bibnamefont{et~al.}, \bibinfo{journal}{Nature Phys.}
  \textbf{\bibinfo{volume}{8}}, \bibinfo{pages}{871–} (\bibinfo{year}{2010}).

\bibitem[{\citenamefont{Comin et~al.}(2014)\citenamefont{Comin, Frano, Yee,
  Yoshida, Eisaki, Schierle, Weschke, Sutarto, He, Soumyanarayanan
  et~al.}}]{comin;s14}
\bibinfo{author}{\bibfnamefont{R.}~\bibnamefont{Comin}},
  \bibinfo{author}{\bibfnamefont{A.}~\bibnamefont{Frano}},
  \bibinfo{author}{\bibfnamefont{M.}~\bibnamefont{Yee}},
  \bibinfo{author}{\bibfnamefont{Y.}~\bibnamefont{Yoshida}},
  \bibinfo{author}{\bibfnamefont{H.}~\bibnamefont{Eisaki}},
  \bibinfo{author}{\bibfnamefont{E.}~\bibnamefont{Schierle}},
  \bibinfo{author}{\bibfnamefont{E.}~\bibnamefont{Weschke}},
  \bibinfo{author}{\bibfnamefont{R.}~\bibnamefont{Sutarto}},
  \bibinfo{author}{\bibfnamefont{F.}~\bibnamefont{He}},
  \bibinfo{author}{\bibfnamefont{A.}~\bibnamefont{Soumyanarayanan}},
  \bibnamefont{et~al.}, \bibinfo{journal}{Science}
  \textbf{\bibinfo{volume}{343}}, \bibinfo{pages}{390} (\bibinfo{year}{2014}).

\bibitem[{\citenamefont{Kivelson et~al.}(2003)\citenamefont{Kivelson, Bindloss,
  Fradkin, Oganesyan, Tranquada, Kapitulnik, and Howald}}]{kivel;rmp03}
\bibinfo{author}{\bibfnamefont{S.~A.} \bibnamefont{Kivelson}},
  \bibinfo{author}{\bibfnamefont{I.~P.} \bibnamefont{Bindloss}},
  \bibinfo{author}{\bibfnamefont{E.}~\bibnamefont{Fradkin}},
  \bibinfo{author}{\bibfnamefont{V.}~\bibnamefont{Oganesyan}},
  \bibinfo{author}{\bibfnamefont{J.~M.} \bibnamefont{Tranquada}},
  \bibinfo{author}{\bibfnamefont{A.}~\bibnamefont{Kapitulnik}},
  \bibnamefont{and} \bibinfo{author}{\bibfnamefont{C.}~\bibnamefont{Howald}},
  \bibinfo{journal}{Rev. Mod. Phys.} \textbf{\bibinfo{volume}{75}},
  \bibinfo{pages}{1201} (\bibinfo{year}{2003}).

\bibitem[{\citenamefont{Billinge and Levin}(2007)}]{billi;s07}
\bibinfo{author}{\bibfnamefont{S.~J.~L.} \bibnamefont{Billinge}}
  \bibnamefont{and} \bibinfo{author}{\bibfnamefont{I.}~\bibnamefont{Levin}},
  \bibinfo{journal}{Science} \textbf{\bibinfo{volume}{316}},
  \bibinfo{pages}{561} (\bibinfo{year}{2007}).

\bibitem[{\citenamefont{Radaelli et~al.}(2002)\citenamefont{Radaelli, Horibe,
  Gutmann, Ishibashi, Chen, Ibberson, Koyama, Hor, Kiryukhin, and
  Cheong}}]{radae;n02}
\bibinfo{author}{\bibfnamefont{P.~G.} \bibnamefont{Radaelli}},
  \bibinfo{author}{\bibfnamefont{Y.}~\bibnamefont{Horibe}},
  \bibinfo{author}{\bibfnamefont{M.~J.} \bibnamefont{Gutmann}},
  \bibinfo{author}{\bibfnamefont{H.}~\bibnamefont{Ishibashi}},
  \bibinfo{author}{\bibfnamefont{C.~H.} \bibnamefont{Chen}},
  \bibinfo{author}{\bibfnamefont{R.~M.} \bibnamefont{Ibberson}},
  \bibinfo{author}{\bibfnamefont{Y.}~\bibnamefont{Koyama}},
  \bibinfo{author}{\bibfnamefont{Y.~S.} \bibnamefont{Hor}},
  \bibinfo{author}{\bibfnamefont{V.}~\bibnamefont{Kiryukhin}},
  \bibnamefont{and} \bibinfo{author}{\bibfnamefont{S.~W.}
  \bibnamefont{Cheong}}, \bibinfo{journal}{Nature}
  \textbf{\bibinfo{volume}{416}}, \bibinfo{pages}{155} (\bibinfo{year}{2002}).

\bibitem[{\citenamefont{Khomskii and Mizokawa}(2005)}]{khoms;prl05}
\bibinfo{author}{\bibfnamefont{D.~I.} \bibnamefont{Khomskii}} \bibnamefont{and}
  \bibinfo{author}{\bibfnamefont{T.}~\bibnamefont{Mizokawa}},
  \bibinfo{journal}{Phys. Rev. Lett.} \textbf{\bibinfo{volume}{94}},
  \bibinfo{pages}{156402} (\bibinfo{year}{2005}).

\bibitem[{\citenamefont{Pascut et~al.}(2014)\citenamefont{Pascut, Haule,
  Gutmann, Barnett, Bombardi, Artyukhin, Birol, Vanderbilt, Yang, Cheong
  et~al.}}]{pascu;prl14}
\bibinfo{author}{\bibfnamefont{G.~L.} \bibnamefont{Pascut}},
  \bibinfo{author}{\bibfnamefont{K.}~\bibnamefont{Haule}},
  \bibinfo{author}{\bibfnamefont{M.~J.} \bibnamefont{Gutmann}},
  \bibinfo{author}{\bibfnamefont{S.~A.} \bibnamefont{Barnett}},
  \bibinfo{author}{\bibfnamefont{A.}~\bibnamefont{Bombardi}},
  \bibinfo{author}{\bibfnamefont{S.}~\bibnamefont{Artyukhin}},
  \bibinfo{author}{\bibfnamefont{T.}~\bibnamefont{Birol}},
  \bibinfo{author}{\bibfnamefont{D.}~\bibnamefont{Vanderbilt}},
  \bibinfo{author}{\bibfnamefont{J.~J.} \bibnamefont{Yang}},
  \bibinfo{author}{\bibfnamefont{S.}~\bibnamefont{Cheong}},
  \bibnamefont{et~al.}, \bibinfo{journal}{Phys. Rev. Lett.}
  \textbf{\bibinfo{volume}{112}}, \bibinfo{pages}{086402}
  (\bibinfo{year}{2014}).

\bibitem[{\citenamefont{Toriyama et~al.}(2014)\citenamefont{Toriyama, Kobori,
  Konishi, Ohta, Sugimoto, Kim, Fujiwara, Pyon, Kudo, and
  Nohara}}]{toriy;jpsj14}
\bibinfo{author}{\bibfnamefont{T.}~\bibnamefont{Toriyama}},
  \bibinfo{author}{\bibfnamefont{M.}~\bibnamefont{Kobori}},
  \bibinfo{author}{\bibfnamefont{T.}~\bibnamefont{Konishi}},
  \bibinfo{author}{\bibfnamefont{Y.}~\bibnamefont{Ohta}},
  \bibinfo{author}{\bibfnamefont{K.}~\bibnamefont{Sugimoto}},
  \bibinfo{author}{\bibfnamefont{J.}~\bibnamefont{Kim}},
  \bibinfo{author}{\bibfnamefont{A.}~\bibnamefont{Fujiwara}},
  \bibinfo{author}{\bibfnamefont{S.}~\bibnamefont{Pyon}},
  \bibinfo{author}{\bibfnamefont{K.}~\bibnamefont{Kudo}}, \bibnamefont{and}
  \bibinfo{author}{\bibfnamefont{M.}~\bibnamefont{Nohara}},
  \bibinfo{journal}{J. Phys. Soc. Jpn} \textbf{\bibinfo{volume}{83}},
  \bibinfo{pages}{033701} (\bibinfo{year}{2014}).

\bibitem[{\citenamefont{Lee et~al.}(2006{\natexlab{b}})\citenamefont{Lee,
  Fujita, {McElroy}, Slezak, Wang, Aiura, Bando, Ishikado, Masui, Zhu
  et~al.}}]{lee;n06}
\bibinfo{author}{\bibfnamefont{J.}~\bibnamefont{Lee}},
  \bibinfo{author}{\bibfnamefont{K.}~\bibnamefont{Fujita}},
  \bibinfo{author}{\bibfnamefont{K.}~\bibnamefont{{McElroy}}},
  \bibinfo{author}{\bibfnamefont{J.~A.} \bibnamefont{Slezak}},
  \bibinfo{author}{\bibfnamefont{M.}~\bibnamefont{Wang}},
  \bibinfo{author}{\bibfnamefont{Y.}~\bibnamefont{Aiura}},
  \bibinfo{author}{\bibfnamefont{H.}~\bibnamefont{Bando}},
  \bibinfo{author}{\bibfnamefont{M.}~\bibnamefont{Ishikado}},
  \bibinfo{author}{\bibfnamefont{T.}~\bibnamefont{Masui}},
  \bibinfo{author}{\bibfnamefont{J.}~\bibnamefont{Zhu}}, \bibnamefont{et~al.},
  \bibinfo{journal}{Nature} \textbf{\bibinfo{volume}{442}},
  \bibinfo{pages}{546} (\bibinfo{year}{2006}{\natexlab{b}}).

\bibitem[{\citenamefont{Bo{\v z}in et~al.}(2007)\citenamefont{Bo{\v z}in,
  Schmidt, {DeConinck}, Paglia, Mitchell, Chatterji, Radaelli, Proffen, and
  Billinge}}]{bozin;prl07}
\bibinfo{author}{\bibfnamefont{E.~S.} \bibnamefont{Bo{\v z}in}},
  \bibinfo{author}{\bibfnamefont{M.}~\bibnamefont{Schmidt}},
  \bibinfo{author}{\bibfnamefont{A.~J.} \bibnamefont{{DeConinck}}},
  \bibinfo{author}{\bibfnamefont{G.}~\bibnamefont{Paglia}},
  \bibinfo{author}{\bibfnamefont{J.~F.} \bibnamefont{Mitchell}},
  \bibinfo{author}{\bibfnamefont{T.}~\bibnamefont{Chatterji}},
  \bibinfo{author}{\bibfnamefont{P.~G.} \bibnamefont{Radaelli}},
  \bibinfo{author}{\bibfnamefont{T.}~\bibnamefont{Proffen}}, \bibnamefont{and}
  \bibinfo{author}{\bibfnamefont{S.~J.~L.} \bibnamefont{Billinge}},
  \bibinfo{journal}{Phys. Rev. Lett.} \textbf{\bibinfo{volume}{98}},
  \bibinfo{pages}{137203} (\bibinfo{year}{2007}).

\bibitem[{\citenamefont{Billinge}(2008)}]{billi;jssc08}
\bibinfo{author}{\bibfnamefont{S.~J.~L.} \bibnamefont{Billinge}},
  \bibinfo{journal}{J. Solid State Chem.} \textbf{\bibinfo{volume}{181}},
  \bibinfo{pages}{1695} (\bibinfo{year}{2008}).

\bibitem[{\citenamefont{Egami and Billinge}(2012)}]{egami;b;utbp12}
\bibinfo{author}{\bibfnamefont{T.}~\bibnamefont{Egami}} \bibnamefont{and}
  \bibinfo{author}{\bibfnamefont{S.~J.~L.} \bibnamefont{Billinge}},
  \emph{\bibinfo{title}{Underneath the Bragg peaks: structural analysis of
  complex materials}} (\bibinfo{publisher}{Elsevier},
  \bibinfo{address}{Amsterdam}, \bibinfo{year}{2012}), \bibinfo{edition}{2nd}
  ed.

\bibitem[{\citenamefont{Yang et~al.}(2012)\citenamefont{Yang, Choi, Oh, Hogan,
  Horibe, Kim, Min, and Cheong}}]{yang;prl12}
\bibinfo{author}{\bibfnamefont{J.~J.} \bibnamefont{Yang}},
  \bibinfo{author}{\bibfnamefont{Y.~J.} \bibnamefont{Choi}},
  \bibinfo{author}{\bibfnamefont{Y.~S.} \bibnamefont{Oh}},
  \bibinfo{author}{\bibfnamefont{A.}~\bibnamefont{Hogan}},
  \bibinfo{author}{\bibfnamefont{Y.}~\bibnamefont{Horibe}},
  \bibinfo{author}{\bibfnamefont{K.}~\bibnamefont{Kim}},
  \bibinfo{author}{\bibfnamefont{B.~I.} \bibnamefont{Min}}, \bibnamefont{and}
  \bibinfo{author}{\bibfnamefont{S.}~\bibnamefont{Cheong}},
  \bibinfo{journal}{Phys. Rev. Lett.} \textbf{\bibinfo{volume}{108}},
  \bibinfo{pages}{116402} (\bibinfo{year}{2012}).

\bibitem[{\citenamefont{Pyon et~al.}(2012)\citenamefont{Pyon, Kudo, and
  Nohara}}]{pyon;jpsj12}
\bibinfo{author}{\bibfnamefont{S.}~\bibnamefont{Pyon}},
  \bibinfo{author}{\bibfnamefont{K.}~\bibnamefont{Kudo}}, \bibnamefont{and}
  \bibinfo{author}{\bibfnamefont{M.}~\bibnamefont{Nohara}},
  \bibinfo{journal}{J. Phys. Soc. Jpn} \textbf{\bibinfo{volume}{81}},
  \bibinfo{pages}{053701} (\bibinfo{year}{2012}).

\bibitem[{\citenamefont{Kamitani et~al.}(2013)\citenamefont{Kamitani, Bahramy,
  Arita, Seki, Arima, Tokura, and Ishiwata}}]{kamit;prb13}
\bibinfo{author}{\bibfnamefont{M.}~\bibnamefont{Kamitani}},
  \bibinfo{author}{\bibfnamefont{M.~S.} \bibnamefont{Bahramy}},
  \bibinfo{author}{\bibfnamefont{R.}~\bibnamefont{Arita}},
  \bibinfo{author}{\bibfnamefont{S.}~\bibnamefont{Seki}},
  \bibinfo{author}{\bibfnamefont{T.}~\bibnamefont{Arima}},
  \bibinfo{author}{\bibfnamefont{Y.}~\bibnamefont{Tokura}}, \bibnamefont{and}
  \bibinfo{author}{\bibfnamefont{S.}~\bibnamefont{Ishiwata}},
  \bibinfo{journal}{Phys. Rev. B} \textbf{\bibinfo{volume}{87}},
  \bibinfo{pages}{180501} (\bibinfo{year}{2013}).

\bibitem[{\citenamefont{Kudo et~al.}(2013)\citenamefont{Kudo, Kobayashi, Pyon,
  and Nohara}}]{kudo;jpsj13}
\bibinfo{author}{\bibfnamefont{K.}~\bibnamefont{Kudo}},
  \bibinfo{author}{\bibfnamefont{M.}~\bibnamefont{Kobayashi}},
  \bibinfo{author}{\bibfnamefont{S.}~\bibnamefont{Pyon}}, \bibnamefont{and}
  \bibinfo{author}{\bibfnamefont{M.}~\bibnamefont{Nohara}},
  \bibinfo{journal}{J. Phys. Soc. Jpn} \textbf{\bibinfo{volume}{82}},
  \bibinfo{pages}{085001} (\bibinfo{year}{2013}).

\bibitem[{\citenamefont{Matsumoto et~al.}(1999)\citenamefont{Matsumoto,
  Taniguchi, Endoh, Takano, and Nagata}}]{matsu;jltp99}
\bibinfo{author}{\bibfnamefont{N.}~\bibnamefont{Matsumoto}},
  \bibinfo{author}{\bibfnamefont{K.}~\bibnamefont{Taniguchi}},
  \bibinfo{author}{\bibfnamefont{R.}~\bibnamefont{Endoh}},
  \bibinfo{author}{\bibfnamefont{H.}~\bibnamefont{Takano}}, \bibnamefont{and}
  \bibinfo{author}{\bibfnamefont{S.}~\bibnamefont{Nagata}},
  \bibinfo{journal}{J. Low Temp. Phys.} \textbf{\bibinfo{volume}{117}},
  \bibinfo{pages}{1129} (\bibinfo{year}{1999}).

\bibitem[{\citenamefont{Morosan et~al.}(2006)\citenamefont{Morosan, Zandbergen,
  Dennis, Bos, Onose, Klimczuk, Ramirez, Ong, and Cava}}]{moros;np06}
\bibinfo{author}{\bibfnamefont{E.}~\bibnamefont{Morosan}},
  \bibinfo{author}{\bibfnamefont{H.~W.} \bibnamefont{Zandbergen}},
  \bibinfo{author}{\bibfnamefont{B.~S.} \bibnamefont{Dennis}},
  \bibinfo{author}{\bibfnamefont{J.~W.~G.} \bibnamefont{Bos}},
  \bibinfo{author}{\bibfnamefont{Y.}~\bibnamefont{Onose}},
  \bibinfo{author}{\bibfnamefont{T.}~\bibnamefont{Klimczuk}},
  \bibinfo{author}{\bibfnamefont{A.~P.} \bibnamefont{Ramirez}},
  \bibinfo{author}{\bibfnamefont{N.~P.} \bibnamefont{Ong}}, \bibnamefont{and}
  \bibinfo{author}{\bibfnamefont{R.~J.} \bibnamefont{Cava}},
  \bibinfo{journal}{Nature Phys.} \textbf{\bibinfo{volume}{2}},
  \bibinfo{pages}{544} (\bibinfo{year}{2006}).

\bibitem[{\citenamefont{Sipos et~al.}(2008)\citenamefont{Sipos, Kusmartseva,
  Akrap, Berger, Forro, and Tutis}}]{sipos;nm08}
\bibinfo{author}{\bibfnamefont{B.}~\bibnamefont{Sipos}},
  \bibinfo{author}{\bibfnamefont{A.~F.} \bibnamefont{Kusmartseva}},
  \bibinfo{author}{\bibfnamefont{A.}~\bibnamefont{Akrap}},
  \bibinfo{author}{\bibfnamefont{H.}~\bibnamefont{Berger}},
  \bibinfo{author}{\bibfnamefont{L.}~\bibnamefont{Forro}}, \bibnamefont{and}
  \bibinfo{author}{\bibfnamefont{E.}~\bibnamefont{Tutis}},
  \bibinfo{journal}{Nature Mater.} \textbf{\bibinfo{volume}{7}},
  \bibinfo{pages}{960} (\bibinfo{year}{2008}).

\bibitem[{\citenamefont{Kusmartseva et~al.}(2009)\citenamefont{Kusmartseva,
  Sipos, Berger, Forro, and Tutis}}]{kusma;prl09}
\bibinfo{author}{\bibfnamefont{A.~F.} \bibnamefont{Kusmartseva}},
  \bibinfo{author}{\bibfnamefont{B.}~\bibnamefont{Sipos}},
  \bibinfo{author}{\bibfnamefont{H.}~\bibnamefont{Berger}},
  \bibinfo{author}{\bibfnamefont{L.}~\bibnamefont{Forro}}, \bibnamefont{and}
  \bibinfo{author}{\bibfnamefont{E.}~\bibnamefont{Tutis}},
  \bibinfo{journal}{Phys. Rev. Lett.} \textbf{\bibinfo{volume}{103}},
  \bibinfo{pages}{236401} (\bibinfo{year}{2009}).

\bibitem[{\citenamefont{Guguchia et~al.}(2017)\citenamefont{Guguchia, {von
  Rohr}, Shermadini, Lee, Banerjee, Wieteska, Marianetti, Frandsen, Luetkens,
  Gong et~al.}}]{guguc;nc17}
\bibinfo{author}{\bibfnamefont{Z.}~\bibnamefont{Guguchia}},
  \bibinfo{author}{\bibfnamefont{F.}~\bibnamefont{{von Rohr}}},
  \bibinfo{author}{\bibfnamefont{Z.}~\bibnamefont{Shermadini}},
  \bibinfo{author}{\bibfnamefont{A.~T.} \bibnamefont{Lee}},
  \bibinfo{author}{\bibfnamefont{S.}~\bibnamefont{Banerjee}},
  \bibinfo{author}{\bibfnamefont{A.~R.} \bibnamefont{Wieteska}},
  \bibinfo{author}{\bibfnamefont{C.~A.} \bibnamefont{Marianetti}},
  \bibinfo{author}{\bibfnamefont{B.~A.} \bibnamefont{Frandsen}},
  \bibinfo{author}{\bibfnamefont{H.}~\bibnamefont{Luetkens}},
  \bibinfo{author}{\bibfnamefont{Z.}~\bibnamefont{Gong}}, \bibnamefont{et~al.},
  \bibinfo{journal}{Nat. Commun.} \textbf{\bibinfo{volume}{8}},
  \bibinfo{pages}{1082} (\bibinfo{year}{2017}).

\bibitem[{\citenamefont{Zhu et~al.}(2016)\citenamefont{Zhu, Ning, Li, Ling,
  Zhang, Zhang, Wang, Liu, Pi, Ma et~al.}}]{zhu;sr16}
\bibinfo{author}{\bibfnamefont{X.}~\bibnamefont{Zhu}},
  \bibinfo{author}{\bibfnamefont{W.}~\bibnamefont{Ning}},
  \bibinfo{author}{\bibfnamefont{L.}~\bibnamefont{Li}},
  \bibinfo{author}{\bibfnamefont{L.}~\bibnamefont{Ling}},
  \bibinfo{author}{\bibfnamefont{R.}~\bibnamefont{Zhang}},
  \bibinfo{author}{\bibfnamefont{J.}~\bibnamefont{Zhang}},
  \bibinfo{author}{\bibfnamefont{K.}~\bibnamefont{Wang}},
  \bibinfo{author}{\bibfnamefont{Y.}~\bibnamefont{Liu}},
  \bibinfo{author}{\bibfnamefont{L.}~\bibnamefont{Pi}},
  \bibinfo{author}{\bibfnamefont{Y.}~\bibnamefont{Ma}}, \bibnamefont{et~al.},
  \bibinfo{journal}{Sci. Rep.} \textbf{\bibinfo{volume}{6}},
  \bibinfo{pages}{26974} (\bibinfo{year}{2016}).

\bibitem[{\citenamefont{Li et~al.}(2017)\citenamefont{Li, Deng, Wang, Liu,
  Abeykoon, Dooryhee, Tomic, Huang, Warren, Bozin et~al.}}]{li;npjqm17}
\bibinfo{author}{\bibfnamefont{L.}~\bibnamefont{Li}},
  \bibinfo{author}{\bibfnamefont{X.}~\bibnamefont{Deng}},
  \bibinfo{author}{\bibfnamefont{Z.}~\bibnamefont{Wang}},
  \bibinfo{author}{\bibfnamefont{Y.}~\bibnamefont{Liu}},
  \bibinfo{author}{\bibfnamefont{M.}~\bibnamefont{Abeykoon}},
  \bibinfo{author}{\bibfnamefont{E.}~\bibnamefont{Dooryhee}},
  \bibinfo{author}{\bibfnamefont{A.}~\bibnamefont{Tomic}},
  \bibinfo{author}{\bibfnamefont{Y.}~\bibnamefont{Huang}},
  \bibinfo{author}{\bibfnamefont{J.~B.} \bibnamefont{Warren}},
  \bibinfo{author}{\bibfnamefont{E.~S.} \bibnamefont{Bozin}},
  \bibnamefont{et~al.}, \bibinfo{journal}{npj Quantum Materials}
  \textbf{\bibinfo{volume}{2}}, \bibinfo{pages}{11} (\bibinfo{year}{2017}).

\bibitem[{\citenamefont{Barath et~al.}(2008)\citenamefont{Barath, Kim, Karpus,
  Cooper, Abbamonte, Fradkin, Morosan, and Cava}}]{barat;prl08}
\bibinfo{author}{\bibfnamefont{H.}~\bibnamefont{Barath}},
  \bibinfo{author}{\bibfnamefont{M.}~\bibnamefont{Kim}},
  \bibinfo{author}{\bibfnamefont{J.~F.} \bibnamefont{Karpus}},
  \bibinfo{author}{\bibfnamefont{S.~L.} \bibnamefont{Cooper}},
  \bibinfo{author}{\bibfnamefont{P.}~\bibnamefont{Abbamonte}},
  \bibinfo{author}{\bibfnamefont{E.}~\bibnamefont{Fradkin}},
  \bibinfo{author}{\bibfnamefont{E.}~\bibnamefont{Morosan}}, \bibnamefont{and}
  \bibinfo{author}{\bibfnamefont{R.~J.} \bibnamefont{Cava}},
  \bibinfo{journal}{Phys. Rev. Lett.} \textbf{\bibinfo{volume}{100}},
  \bibinfo{pages}{106402} (\bibinfo{year}{2008}).

\bibitem[{\citenamefont{{Castro Neto}}(2001)}]{castr;prl01}
\bibinfo{author}{\bibfnamefont{A.~H.} \bibnamefont{{Castro Neto}}},
  \bibinfo{journal}{Phys. Rev. Lett.} \textbf{\bibinfo{volume}{86}},
  \bibinfo{pages}{4382} (\bibinfo{year}{2001}).

\bibitem[{\citenamefont{Dai et~al.}(2014)\citenamefont{Dai, Haule, Yang, Oh,
  Cheong, and Wu}}]{dai;prb14}
\bibinfo{author}{\bibfnamefont{J.}~\bibnamefont{Dai}},
  \bibinfo{author}{\bibfnamefont{K.}~\bibnamefont{Haule}},
  \bibinfo{author}{\bibfnamefont{J.~J.} \bibnamefont{Yang}},
  \bibinfo{author}{\bibfnamefont{Y.~S.} \bibnamefont{Oh}},
  \bibinfo{author}{\bibfnamefont{S.}~\bibnamefont{Cheong}}, \bibnamefont{and}
  \bibinfo{author}{\bibfnamefont{W.}~\bibnamefont{Wu}}, \bibinfo{journal}{Phys.
  Rev. B} \textbf{\bibinfo{volume}{90}}, \bibinfo{pages}{235121}
  (\bibinfo{year}{2014}).

\bibitem[{\citenamefont{Oh et~al.}(2013)\citenamefont{Oh, Yang, Horibe, and
  Cheong}}]{oh;prl13}
\bibinfo{author}{\bibfnamefont{Y.~S.} \bibnamefont{Oh}},
  \bibinfo{author}{\bibfnamefont{J.~J.} \bibnamefont{Yang}},
  \bibinfo{author}{\bibfnamefont{Y.}~\bibnamefont{Horibe}}, \bibnamefont{and}
  \bibinfo{author}{\bibfnamefont{S.}~\bibnamefont{Cheong}},
  \bibinfo{journal}{Phys. Rev. Lett.} \textbf{\bibinfo{volume}{110}},
  \bibinfo{pages}{127209} (\bibinfo{year}{2013}).

\bibitem[{\citenamefont{Kim et~al.}(2015)\citenamefont{Kim, Kim, Ko, Lee, Park,
  Yang, Cheong, and Min}}]{kim;prl15}
\bibinfo{author}{\bibfnamefont{K.}~\bibnamefont{Kim}},
  \bibinfo{author}{\bibfnamefont{S.}~\bibnamefont{Kim}},
  \bibinfo{author}{\bibfnamefont{K.}~\bibnamefont{Ko}},
  \bibinfo{author}{\bibfnamefont{H.}~\bibnamefont{Lee}},
  \bibinfo{author}{\bibfnamefont{J.}~\bibnamefont{Park}},
  \bibinfo{author}{\bibfnamefont{J.~J.} \bibnamefont{Yang}},
  \bibinfo{author}{\bibfnamefont{S.}~\bibnamefont{Cheong}}, \bibnamefont{and}
  \bibinfo{author}{\bibfnamefont{B.~I.} \bibnamefont{Min}},
  \bibinfo{journal}{Phys. Rev. Lett.} \textbf{\bibinfo{volume}{114}},
  \bibinfo{pages}{136401} (\bibinfo{year}{2015}).

\bibitem[{\citenamefont{Ivashko et~al.}(2017)\citenamefont{Ivashko, Yang,
  Destraz, Martino, Chen, Guo, Yuan, Pisoni, Matus, Pyon et~al.}}]{ivash;sr17}
\bibinfo{author}{\bibfnamefont{O.}~\bibnamefont{Ivashko}},
  \bibinfo{author}{\bibfnamefont{L.}~\bibnamefont{Yang}},
  \bibinfo{author}{\bibfnamefont{D.}~\bibnamefont{Destraz}},
  \bibinfo{author}{\bibfnamefont{E.}~\bibnamefont{Martino}},
  \bibinfo{author}{\bibfnamefont{Y.}~\bibnamefont{Chen}},
  \bibinfo{author}{\bibfnamefont{C.~Y.} \bibnamefont{Guo}},
  \bibinfo{author}{\bibfnamefont{H.~Q.} \bibnamefont{Yuan}},
  \bibinfo{author}{\bibfnamefont{A.}~\bibnamefont{Pisoni}},
  \bibinfo{author}{\bibfnamefont{P.}~\bibnamefont{Matus}},
  \bibinfo{author}{\bibfnamefont{S.}~\bibnamefont{Pyon}}, \bibnamefont{et~al.},
  \bibinfo{journal}{Sci. Rep.} \textbf{\bibinfo{volume}{7}},
  \bibinfo{pages}{17157} (\bibinfo{year}{2017}).

\bibitem[{\citenamefont{Joseph et~al.}(2013)\citenamefont{Joseph, Bendele,
  Simonelli, Maugeri, Pyon, Kudo, Nohara, Mizokawa, and Saini}}]{josep;prb13}
\bibinfo{author}{\bibfnamefont{B.}~\bibnamefont{Joseph}},
  \bibinfo{author}{\bibfnamefont{M.}~\bibnamefont{Bendele}},
  \bibinfo{author}{\bibfnamefont{L.}~\bibnamefont{Simonelli}},
  \bibinfo{author}{\bibfnamefont{L.}~\bibnamefont{Maugeri}},
  \bibinfo{author}{\bibfnamefont{S.}~\bibnamefont{Pyon}},
  \bibinfo{author}{\bibfnamefont{K.}~\bibnamefont{Kudo}},
  \bibinfo{author}{\bibfnamefont{M.}~\bibnamefont{Nohara}},
  \bibinfo{author}{\bibfnamefont{T.}~\bibnamefont{Mizokawa}}, \bibnamefont{and}
  \bibinfo{author}{\bibfnamefont{N.~L.} \bibnamefont{Saini}},
  \bibinfo{journal}{Phys. Rev. B} \textbf{\bibinfo{volume}{88}},
  \bibinfo{pages}{224109} (\bibinfo{year}{2013}).

\bibitem[{\citenamefont{Paris et~al.}(2016)\citenamefont{Paris, Joseph,
  Iadecola, Marini, Ishii, Kudo, Pascarelli, Nohara, Mizokawa, and
  Saini}}]{paris;prb16}
\bibinfo{author}{\bibfnamefont{E.}~\bibnamefont{Paris}},
  \bibinfo{author}{\bibfnamefont{B.}~\bibnamefont{Joseph}},
  \bibinfo{author}{\bibfnamefont{A.}~\bibnamefont{Iadecola}},
  \bibinfo{author}{\bibfnamefont{C.}~\bibnamefont{Marini}},
  \bibinfo{author}{\bibfnamefont{H.}~\bibnamefont{Ishii}},
  \bibinfo{author}{\bibfnamefont{K.}~\bibnamefont{Kudo}},
  \bibinfo{author}{\bibfnamefont{S.}~\bibnamefont{Pascarelli}},
  \bibinfo{author}{\bibfnamefont{M.}~\bibnamefont{Nohara}},
  \bibinfo{author}{\bibfnamefont{T.}~\bibnamefont{Mizokawa}}, \bibnamefont{and}
  \bibinfo{author}{\bibfnamefont{N.~L.} \bibnamefont{Saini}},
  \bibinfo{journal}{Phys. Rev. B} \textbf{\bibinfo{volume}{93}},
  \bibinfo{pages}{134109} (\bibinfo{year}{2016}).

\bibitem[{\citenamefont{Bo\v{z}in et~al.}(2011)\citenamefont{Bo\v{z}in,
  Masadeh, Hor, Mitchell, and Billinge}}]{bozin;prl11}
\bibinfo{author}{\bibfnamefont{E.~S.} \bibnamefont{Bo\v{z}in}},
  \bibinfo{author}{\bibfnamefont{A.~S.} \bibnamefont{Masadeh}},
  \bibinfo{author}{\bibfnamefont{Y.~S.} \bibnamefont{Hor}},
  \bibinfo{author}{\bibfnamefont{J.~F.} \bibnamefont{Mitchell}},
  \bibnamefont{and} \bibinfo{author}{\bibfnamefont{S.~J.~L.}
  \bibnamefont{Billinge}}, \bibinfo{journal}{Phys. Rev. Lett.}
  \textbf{\bibinfo{volume}{106}}, \bibinfo{pages}{045501}
  (\bibinfo{year}{2011}).

\bibitem[{\citenamefont{Bo\v{z}in et~al.}(2014)\citenamefont{Bo\v{z}in, Knox,
  Juh\'{a}s, Hor, Mitchell, and Billinge}}]{bozin;sr14}
\bibinfo{author}{\bibfnamefont{E.~S.} \bibnamefont{Bo\v{z}in}},
  \bibinfo{author}{\bibfnamefont{K.~R.} \bibnamefont{Knox}},
  \bibinfo{author}{\bibfnamefont{P.}~\bibnamefont{Juh\'{a}s}},
  \bibinfo{author}{\bibfnamefont{Y.~S.} \bibnamefont{Hor}},
  \bibinfo{author}{\bibfnamefont{J.~F.} \bibnamefont{Mitchell}},
  \bibnamefont{and} \bibinfo{author}{\bibfnamefont{S.~J.~L.}
  \bibnamefont{Billinge}}, \bibinfo{journal}{Sci. Rep.}
  \textbf{\bibinfo{volume}{4}}, \bibinfo{pages}{4081} (\bibinfo{year}{2014}).

\bibitem[{\citenamefont{Cao et~al.}(2013)\citenamefont{Cao, Chakoumakos, Chen,
  Yan, {McGuire}, Yang, Custelcean, Zhou, Singh, and Mandrus}}]{cao;prb13}
\bibinfo{author}{\bibfnamefont{H.}~\bibnamefont{Cao}},
  \bibinfo{author}{\bibfnamefont{B.~C.} \bibnamefont{Chakoumakos}},
  \bibinfo{author}{\bibfnamefont{X.}~\bibnamefont{Chen}},
  \bibinfo{author}{\bibfnamefont{J.}~\bibnamefont{Yan}},
  \bibinfo{author}{\bibfnamefont{M.~A.} \bibnamefont{{McGuire}}},
  \bibinfo{author}{\bibfnamefont{H.}~\bibnamefont{Yang}},
  \bibinfo{author}{\bibfnamefont{R.}~\bibnamefont{Custelcean}},
  \bibinfo{author}{\bibfnamefont{H.~D.} \bibnamefont{Zhou}},
  \bibinfo{author}{\bibfnamefont{D.~J.} \bibnamefont{Singh}}, \bibnamefont{and}
  \bibinfo{author}{\bibfnamefont{D.}~\bibnamefont{Mandrus}},
  \bibinfo{journal}{Phys. Rev. B} \textbf{\bibinfo{volume}{88}},
  \bibinfo{pages}{115122} (\bibinfo{year}{2013}).

\bibitem[{\citenamefont{Lazarevic et~al.}(2014)\citenamefont{Lazarevic, Bozin,
  Scepanovic, Opacic, Lei, Petrovic, and Popovic}}]{lazar;prb14}
\bibinfo{author}{\bibfnamefont{N.}~\bibnamefont{Lazarevic}},
  \bibinfo{author}{\bibfnamefont{E.~S.} \bibnamefont{Bozin}},
  \bibinfo{author}{\bibfnamefont{M.}~\bibnamefont{Scepanovic}},
  \bibinfo{author}{\bibfnamefont{M.}~\bibnamefont{Opacic}},
  \bibinfo{author}{\bibfnamefont{H.}~\bibnamefont{Lei}},
  \bibinfo{author}{\bibfnamefont{C.}~\bibnamefont{Petrovic}}, \bibnamefont{and}
  \bibinfo{author}{\bibfnamefont{Z.~V.} \bibnamefont{Popovic}},
  \bibinfo{journal}{Phys. Rev. B} \textbf{\bibinfo{volume}{89}},
  \bibinfo{pages}{224301} (\bibinfo{year}{2014}).

\bibitem[{\citenamefont{Chupas et~al.}(2003)\citenamefont{Chupas, Qiu, Hanson,
  Lee, Grey, and Billinge}}]{chupa;jac03}
\bibinfo{author}{\bibfnamefont{P.~J.} \bibnamefont{Chupas}},
  \bibinfo{author}{\bibfnamefont{X.}~\bibnamefont{Qiu}},
  \bibinfo{author}{\bibfnamefont{J.~C.} \bibnamefont{Hanson}},
  \bibinfo{author}{\bibfnamefont{P.~L.} \bibnamefont{Lee}},
  \bibinfo{author}{\bibfnamefont{C.~P.} \bibnamefont{Grey}}, \bibnamefont{and}
  \bibinfo{author}{\bibfnamefont{S.~J.~L.} \bibnamefont{Billinge}},
  \bibinfo{journal}{J. Appl. Crystallogr.} \textbf{\bibinfo{volume}{36}},
  \bibinfo{pages}{1342} (\bibinfo{year}{2003}).

\bibitem[{\citenamefont{Hammersley et~al.}(1996)\citenamefont{Hammersley,
  Svenson, Hanfland, and Hauserman}}]{hamme;hpr96}
\bibinfo{author}{\bibfnamefont{A.~P.} \bibnamefont{Hammersley}},
  \bibinfo{author}{\bibfnamefont{S.~O.} \bibnamefont{Svenson}},
  \bibinfo{author}{\bibfnamefont{M.}~\bibnamefont{Hanfland}}, \bibnamefont{and}
  \bibinfo{author}{\bibfnamefont{D.}~\bibnamefont{Hauserman}},
  \bibinfo{journal}{High Pressure Res.} \textbf{\bibinfo{volume}{14}},
  \bibinfo{pages}{235} (\bibinfo{year}{1996}).

\bibitem[{\citenamefont{Juh\'{a}s et~al.}(2013)\citenamefont{Juh\'{a}s, Davis,
  Farrow, and Billinge}}]{juhas;jac13}
\bibinfo{author}{\bibfnamefont{P.}~\bibnamefont{Juh\'{a}s}},
  \bibinfo{author}{\bibfnamefont{T.}~\bibnamefont{Davis}},
  \bibinfo{author}{\bibfnamefont{C.~L.} \bibnamefont{Farrow}},
  \bibnamefont{and} \bibinfo{author}{\bibfnamefont{S.~J.~L.}
  \bibnamefont{Billinge}}, \bibinfo{journal}{J. Appl. Crystallogr.}
  \textbf{\bibinfo{volume}{46}}, \bibinfo{pages}{560} (\bibinfo{year}{2013}).

\bibitem[{\citenamefont{Farrow et~al.}(2007)\citenamefont{Farrow, Juh\'as, Liu,
  Bryndin, {Bo\v zin}, Bloch, Proffen, and Billinge}}]{farro;jpcm07}
\bibinfo{author}{\bibfnamefont{C.~L.} \bibnamefont{Farrow}},
  \bibinfo{author}{\bibfnamefont{P.}~\bibnamefont{Juh\'as}},
  \bibinfo{author}{\bibfnamefont{J.}~\bibnamefont{Liu}},
  \bibinfo{author}{\bibfnamefont{D.}~\bibnamefont{Bryndin}},
  \bibinfo{author}{\bibfnamefont{E.~S.} \bibnamefont{{Bo\v zin}}},
  \bibinfo{author}{\bibfnamefont{J.}~\bibnamefont{Bloch}},
  \bibinfo{author}{\bibfnamefont{T.}~\bibnamefont{Proffen}}, \bibnamefont{and}
  \bibinfo{author}{\bibfnamefont{S.~J.~L.} \bibnamefont{Billinge}},
  \bibinfo{journal}{J. Phys: Condens. Mat.} \textbf{\bibinfo{volume}{19}},
  \bibinfo{pages}{335219} (\bibinfo{year}{2007}).

\bibitem[{\citenamefont{Hockings and White}(1960)}]{hocki;jpc60}
\bibinfo{author}{\bibfnamefont{E.~F.} \bibnamefont{Hockings}} \bibnamefont{and}
  \bibinfo{author}{\bibfnamefont{J.~G.} \bibnamefont{White}},
  \bibinfo{journal}{J. Phys. Chem.} \textbf{\bibinfo{volume}{64}},
  \bibinfo{pages}{1042} (\bibinfo{year}{1960}).

\end{thebibliography}
\end{document}